\documentclass[usenatbib]{basi}
\usepackage[T1]{fontenc}
\usepackage[british]{babel}
\usepackage[varg]{txfonts}

\usepackage{rotating}
\usepackage{dcolumn}

\usepackage{url}

\begin{document}

\title[SPAM]{SPAM: A data reduction recipe for high-resolution, low-frequency radio-interferometric observations}

\author[H.~T.~Intema]{H.~T.~Intema$^1$
\thanks{email: \texttt{hintema@nrao.edu}} \\
$^1$National Radio Astronomy Observatory, 1003 Lopezville Road, Socorro, NM 87801-0387, USA}

\pubyear{2014}
\volume{...}
\pagerange{\pageref{firstpage}--\pageref{lastpage}}
\date{Received --- ; accepted ---}

\maketitle


\label{firstpage}

\begin{abstract}
High-resolution astronomical imaging at sub-GHz radio frequencies has been available for more than 15 years, with the VLA at 74 and 330 MHz, and the GMRT at 150, 240, 330 and 610 MHz. Recent developments include wide-bandwidth upgrades for VLA and GMRT, and commissioning of the aperture-array-based, multi-beam telescope LOFAR. A common feature of these telescopes is the necessity to deconvolve the very many detectable sources within their wide fields-of-view and beyond. This is complicated by gain variations in the radio signal path that depend on viewing direction. One such example is phase errors due to the ionosphere.

Here I discuss the inner workings of SPAM, a set of AIPS-based data reduction scripts in Python that includes direction-dependent calibration and imaging. Since its first version in 2008, SPAM has been applied to many GMRT data sets at various frequencies. Many valuable lessons were learned, and translated into various SPAM software modifications. Nowadays, semi-automated SPAM data reduction recipes can be applied to almost any GMRT data set, yielding good quality continuum images comparable with (or often better than) hand-reduced results. SPAM is currently being migrated from AIPS to CASA with an extension to handle wide bandwidths. This is aimed at providing users of the VLA low-band system and the upcoming wide-bandwidth GMRT with the necessary data reduction tools.
\end{abstract}

\begin{keywords}
atmospheric effects -- 
methods: data analysis --
instrumentation: interferometers
\end{keywords}


\section{Introduction}
\label{sec:intro}

Low-frequency radio telescopes have relatively large fields-of-view, because the telescope primary beam size scales with wavelength. Current high-resolution (sub-arcminute) radio interferometers that operate at sub-GHz frequencies, like the Giant Metrewave Radio Telescope (GMRT), the LOw-Frequency ARray (LOFAR), or the Very Large Array (VLA), have field diameters measured in (sometimes tens of) degrees. Typical challenges of handling high-resolution, low-frequency continuum observations are (i) imaging and deconvolving many sources inside (and outside!) the large primary beam area, (ii) direction-dependent (DD) visibility amplitude and phase variations across the field-of-view due to antenna beam patterns, pointing errors, ionosphere, etc., and the abundance of Radio Frequency Interference (RFI).

Processing data coming from these telescopes can be daunting, since all the challenges mentioned above (and more) needs to be dealt with at the same time. In data reduction, the \emph{art} is to determine at any given time which effect is dominant in limiting the image quality. Generally, a few \emph{big effects} stand out and are generally easy to identify and fix. But subsequenty, many \emph{smaller effect} will present themelves, which may be more difficult to recognize, and harder to identify and disentangle. A effective way to mitigate the smaller effects is to perform iterations of (DD) calibration, imaging and flagging, in each step refining the quality of both the visibility data and the reconstructed sky model.


\section{SPAM: an overview}
\label{sec:overview}

Developments towards a semi-automated data reduction package started in 2006, driven by the complicated reduction of GMRT 150 MHz observations on galaxy cluster Abell 2256 (and other similar data sets). Several challenges presented themselves at the same time: (i) the target contains faint, diffuse emission in the presence of bright interfering sources, (ii) the data was strongly affected by RFI on many baselines, (iii) the data was strongly affected by ionospheric phase distortions, (iv) the data was affected by instrumental instabilities. It quickly became clear that this data required an algorithm to correct for ionospheric phase errors, and possibly other DD effects. Simultaneously followed the desire to automate several trivial but very time-consuming data reduction steps. 

For the basis of these new developments a stable data reduction package was sought, with the possibility to easily expand the functionality. This was found in the combination of the data reduction package AIPS
\citep{2003ASSL..285..109G} 
and the powerful high-level programming language Python (and its standard scientific libraries like scipy, pylab, matplotlib, and numpy), with the ParselTongue interface 
\citep{2006ASPC..351..497K} 
providing access to AIPS tasks, data files (images and visibilities) and tables from Python. A new Python module was created named \emph{SPAM}, an acronym for \emph{Source Peeling and Atmospheric Modeling}, encapsulating high-level data reduction functions and new algorithm development. 

SPAM-based data reductions, i.e. for GMRT observations, have been captured in Python scripts that execute AIPS tasks directly (mostly during initial reduction steps), call high-level functions that encompass multiple AIPS /ParselTongue calls, and require few manual operations. These scripted data reductions automatically keep a history of the data reduction steps, and provide a standard that is well-tested and reproducible. At the same time, the SPAM user is required to perform some manual steps, and can easily decide to change the order of data reduction steps depending on the target field geometry and the intended science, therefore the term data reduction \emph{recipe} is more appropriate than \emph{pipeline}. Note that SPAM assumes an unpolarized sky, observed with limited ($\lesssim 20$~percent) fractional bandwidth.

SPAM recipes are available for all sub-GHz GMRT frequencies, developed and refined over the last several years, with minimal differences between frequencies. The general structure of this data reduction is described in 
\citet{2009A&A...501.1185I}, 
and includes \emph{measuring} DD ionospheric phase errors through peeling of bright sources within the science field 
\citep[e.g.,][]{2004SPIE.5489..817N},  
\emph{modeling} these errors with a single- or multi-layer ionospheric phase model, and \emph{applying} ionospheric phase corrections during wide-field (re-)imaging of the science field (see Section~\ref{sec:ioncal}). 

A combination of RFI- and bad data mitigation routines are used inbetween rounds of calibration and imaging. \emph{Classical outlier removal} excises visibilities mostly based on excessive visibility amplitudes and statistical outlier rejection along the time and frequency axes. RFI \emph{subtraction} models and subtracts low-level, quasi-continuous, ground-based RFI based on its fringe-rotation signature 
\citep[e.g.,][]{2009ApJ...696..885A}. 
And \emph{ripple killing}  excises bad visibility data based on a combination of high visibility amplitudes, high visibility weights and high imaging weights (i.e., though uniform weighting). This is done by Fourier-transforming residual images back onto the UV-plane, identifying high-amplitude UV-cells, and rejecting all visibilities that fall in these cells\footnote{E.g., see the AIPS tutorial by En{\ss}lin \& Kronberg \\
\url{http://www.mpa-garching.mpg.de/~ensslin/Paper/Reduce/remove.ps.gz}}.
A more elaborate description will be given in an upcoming paper (Intema et al., in preparation).


\subsection{SPAM ionospheric calibration and imaging}
\label{sec:ioncal}

Following 
\citet{2005ASPC..345..399L} 
there are four ionospheric calibration regimes, depending on the size of the radio interferometer array and the size of the field-of-view, both relative to the scale of phase structure in the ionosphere. This makes most sense when imagining a single horizontal density wave (or a bubble) in a thin layer at a fixed height. Regime~4 is the most complex case, when both the horizontal size of the array and the projected field-of-view at ionospheric height are similar or larger than the horizontal scale of the ionospheric phase structure. In this regime, when looking towards a single source in the field-of-view, the higher-order phase structure over the array will cause both an apparent position shift and a source deformation. When looking at a second source in the same field-of-view at sufficient distance from the first, both the apparent position shift and the source distortion will be different. 

The SPAM ionospheric calibration strategy is designed to operate in regime~4, meaning that it will also work in the less complex regimes 1--3. In an automated peeling routine, measurements of the ionospheric phase structure are obtained by phase calibrating on (typically 10--20) bright sources within the field-of-view. Per time interval (typically 10--20~seconds), the measured phases of all source--antenna pairs are mapped onto a common domain along their line-of-sight, i.e., at the pierce points through a virtual phase screen at fixed height. The  phases are fit with an optimized set of base functions, reproducing the measured phases, and predicting DD phases corrections in arbitrary viewing directions during imaging. 

The automated wide-field imager applies any relevant DD phase corrections on the fly in a modified version of polyhedron (facet-based) imaging and Cotton-Schwab CLEAN deconvolution
\citep[e.g., see][]{1999ASPC..180.....T}. 
The intrinsic book-keeping issues of faceted imaging are largely hidden from the user. The imager also includes an iterative, robust clean boxing scheme, and simultaneous imaging of the primary beam area and bright outlier fields. In principle, the generic imager can handle other DD effects as well, like asymmetric primary beams or pointing errors, as long as a model is available to generate the DD correction tables. 


\section{Summary}
\label{sec:summary}

SPAM is an AIPS-based Python package that provides semi-automated data reduction scripts for all sub-GHz frequencies at GMRT. Apart from well-tested standard data reduction steps, SPAM includes direction-dependent (ionospheric) calibration and imaging, and alternative RFI- and bad data mitigation methods. With SPAM, GMRT data reductions are highly efficient, highly reproducible, and give good quality results. SPAM is available for everyone to use\footnote{\url{https://safe.nrao.edu/wiki/bin/view/Main/HuibIntemaSpam}}.

SPAM for CASA, currently under development, will also provide wide-bandwidth ionospheric calibration, direction-dependent ionospheric Faraday rotation calibration, and will naturally benefit from the advanced modes (and continuous improvements) of the CASA imager
\citep[e.g., see][and references therein]{2013ApJ...770...91B}. 

%
%


%




%


\label{lastpage}
\end{document}